\documentclass[preprint]{elsarticle}

\usepackage{lipsum}
\usepackage{lineno}
\usepackage{hyperref}
\usepackage{inputenc}
\usepackage{graphicx}
\usepackage{amsmath}
\usepackage[official]{eurosym}

\journal{Computers in Biology and Medicine}

\begin{document}

\begin{frontmatter}
\title{Non-invasive estimation of left ventricle elastance using a multi-compartment lumped parameter model and gradient-based optimization with forward-mode automatic differentiation}
\author{Ryno Laubscher and Johan van der Merwe}
\address{Institute for Biomedical Engineering, Department of Mechanical and Mechatronic Engineering, Stellenbosch University, Stellenbosch, South Africa}
\author{Jacques Liebenberg and Philip Herbst}
\address{Division of Cardiology, Faculty of Medicine and Health Sciences, Stellenbosch University, Cape Town, South Africa}

\begin{abstract}
Accurate estimates of left ventricle elastances based on non-invasive measurements are required for clinical decision-making during treatment of valvular diseases. The present study proposes a parameter discovery approach based on a lumped parameter model of the cardiovascular system in conjunction with optimization and non-invasive, clinical input measurements to approximate important cardiac parameters, including left ventricle elastances. A subset of parameters of a multi-compartment lumped parameter model was estimated using 1st order Adam gradient descent and hybrid Adam and 2nd order quasi-Newton limited-memory Broyden-Fletcher-Goldfarb-Shanno optimization routines. Forward-mode automatic differentiation was used to estimate the Jacobian matrices and compared to the common finite differences approach. Synthetic data of healthy and diseased hearts were generated as proxies for non-invasive clinical measurements and used to evaluate the algorithm. Twelve parameters including left ventricle elastances were selected for optimization based on 99\% explained variation in mean left ventricle pressure and volume. The hybrid optimization strategy yielded the best overall results compared to 1st order optimization with automatic differentiation and finite difference approaches, with mean absolute percentage errors ranging from 6.67\% to 14,14\%. Errors in left ventricle elastance estimates for simulated aortic stenosis and mitral regurgitation were smallest when including synthetic measurements for arterial pressure and valvular flow rate at approximately 2\% and degraded to roughly 5\% when including volume trends as well. However, the latter resulted in better tracking of the left ventricle pressure waveforms and may be considered when the necessary equipment is available.
\end{abstract}


\end{frontmatter}

\section{Introduction}

The rapid rise of cardiovascular diseases (CVDs) in sub-Saharan Africa (SSA) along with the high rates of communicable diseases in these regions, results in a double burden experienced by the public health care sector, undermining its effectiveness \cite{BeLue2009}. CVD related deaths in SSA account for approximately 11\% of all deaths in the region \cite{Keates2017}. The most prevalent aetiologies of heart failure in SSA that eventually result in morbidity and mortality are rheumatic heart disease (RHD), ischemic heart disease and hypertensive heart disease \cite{Yuyun2020}. RHD is especially common in SSA as a result of low income status, poor living conditions and limited access to medical care. The high prevalence and late detection results in region-specific deaths that accounts for 23\% of the world’s deaths related to this disease. RHD usually involves the aortic or mitral valve, resulting in either valvular stenosis or regurgitation. Long standing neglected severe valve lesions ultimately leads to cardiac decompensation and impairs the hearts ability to deliver oxygen rich blood to various organs.  Assessment of intrinsic ventricular function is of critical importance in the management of patients with valve disease and is an important trigger for recommending valve intervention, including valve replacement surgery \cite{Serbia2021}. Unfortunately, the majority of clinically accessible parameters used to assess ventricular function are measures of pump function that are significantly load dependent, the most important of these, left ventricular ejection fraction (LVEF). In the presence of severe mitral regurgitation the intrinsic ventricular muscle function is difficult to judge using LVEF as the LV offloads into a lower pressure left atrium (LA) complicating its assessment. In severe aortic stenosis, intrinsic ventricular muscle dysfunction from a number of associated comorbid conditions, can similarly  be difficult to differentiate from excessively high afterload reducing muscle shortening. LV elastance, a load independent parameter, is an appealing alternative and is considered the gold standard for LV systolic function assessment and ventricular – arterial interaction \cite{Gayat2011}. To accurately determine the LV elastance for a patient, invasive, multielectrode conductance catheterisation is typically utilised \cite{Kass1986}. In developing countries, this invasive procedure is not readily available to clinicians, is expensive and is not without risk. To non-invasively approximate a patient’s LV elastance, clinicians typically use systolic and diastolic blood pressures, Doppler-based measured stroke volume and derived LV ejection fraction with Chen’s formula \cite{Chen2001}. Using Chen’s formula with abnormal ventricular loading conditions, such as those observed in patients with valvular diseases, leads to large variability in the accuracy of the calculated LV elastance due to the load – dependant nature of LVEF used in this calculation \cite{Bikia2021}. Therefore, a need still exists to develop a robust approach to non-invasively approximate LV elastances for diseased heart valves and ventricles.

An exciting, but experimental, approach to achieve robust non-invasive estimation of patient LV elastance is the combination of cardiovascular lumped parameter modelling (LPM), optimisation algorithms, and clinical measurements. In this approach, the LPM acts as a mechanistic bridge between the non-invasive measurements and the desired parameters (LV elastance). The optimisation algorithms are used to minimise the difference between the LPM results and the available measurements by tuning the mathematical model parameters such as LV elastance, arterial resistance and pulmonary artery compliance.

Recently, several researchers have investigated the feasibility of using LPMs in conjunction with optimisation algorithms to find unknown cardiovascular parameters. Herewith follows a brief overview of selected research papers on this subject. Huang and Ying \cite{Huang2020} developed an on-line parameter identification algorithm to estimate the parameters of a 5 component arterial model. The authors used synthetic data as a proxy for clinical measurements. The synthetic data was generated using a multi-compartment LPM of a complete cardiovascular system. To effectively approximate the parameters, the mean squared error between the transient synthetic data and the corresponding model predictions was minimized using a MATLAB Simulink model along with its accompanying non-linear optimization package (fmincon \cite{Mathworks2022}). The developed algorithm was capable of recovering the unknown arterial resistances and compliances from the synthetic data. Zhang et al. \cite{Zhang2020a} developed a 0D-1D model of a complete cardiovascular system and used clinically measured data to identify key parameters such as arterial stiffnesses, peak flow time, distal end scale factors and resistances. Non-invasive patient measurements were performed to estimate artery diameters, ventricular stroke volumes, blood pressure waveforms and axial blood flow velocities. To find the unknown parameter set, the mean squared differences between the model predictions and measured brachial arterial pressures and aortic velocities were minimized using the Levenberg-Marquardt optimization algorithm. Seeing as only non-invasive stroke volumes and brachial pressures were used by the authors, it limits the approach to persons with no heart diseases such as aortic valve stenosis. It should be stated that modelling heart diseases was not the purpose of the previous authors’ research. Colunga et al. \cite{Colunga2020} used invasive right heart catheterisation data and LPMs to identify cardiovascular parameters such as pulmonary and systemic resistances, pulmonary valve resistance, pulmonary and systemic arterial compliances and right ventricle elastances. The LPM developed by the authors consisted of 6 components (e.g. systemic artery, left ventricle, pulmonary artery) and the squared differences between the predicted and clinically measured arterial blood pressures, right heart ventricle pressures and cardiac outputs were minimised using the Levenberg-Marquardt optimisation routine in MATLAB for various patients. The optimised LPMs could accurately recreate measured waveforms but seeing as the identified parameters were not directly measured, the absolute accuracy of the identified parameters is unknown. Keshavarz-Motamed \cite{Keshavarz-Motamed2020} developed a comprehensive workflow coined C3VI-CMF, to identify circulatory system parameters and heart functions using MATLAB Simscape and the non-linear design optimisation toolbox in MATLAB Simulink. The developed tool uses Doppler echography and arm cuff pressure measurements along with a multi-compartment LPM to identify aortic compliance, systemic arterial and venous compliance, systemic arterial resistance and mean pulmonary valve flow rate. The remainder of the LPM parameters, such as ventricular elastances, were set to typical clinical values. The ventricular pressure waveforms predicted by the LPM with the identified parameters were compared to the ventricular catheterisation data and showed excellent agreement. In \cite{Keshavarz-Motamed2020}, errors between the predicted LPM outputs and the non-invasive measurements were optimised using the fmincon function in MATLAB. Bjørdalsbakke et al. \cite{Bjordalsbakke2022} developed a program capable of estimating systemic loop parameters such as systemic compliances, LV elastances (diastolic and systolic), systemic resistance and aorta blood inertia using the trust region reflective algorithm \cite{Branch1999} from the free and opensource Python library SciPy. The developed LPM did not consider the pulmonary loop and approximated the systemic loop using a 3-element component. The authors developed an innovative technique called the stepwise subset reduction method to reduce the effect of less important parameters during optimisation. Synthetic data was generated using the LPM model and the difference between the model predictions and the synthetic data was used to identify the unknown parameters. The achieved mean absolute percentage errors (MAPEs) ranged between 1-10\% for the identified parameters.

For the previously discussed MATLAB- and Python-based models, the parameter-cost function Jacobian matrices used in the optimisation algorithms are estimated using the finite differences approach \cite{Mathworks2022}, \cite{SciPyOrganisation2021}.  Gradients obtained using finite differences are typically computationally expensive for large problems, such as system-level transient simulations, and tend to be numerically unstable seeing as it only approximates the gradients \cite{Marsden2014}. An alternative approach to estimate the Jacobian matrices is to use a method typically applied in deep learning called automatic differentiation (autodiff) \cite{Kochenderfer2019a}, which calculates the analytical gradients using the chain rule and computational graphs constructed from the mathematical operations in the computer model. In the present work, we propose a parameter discovery algorithm which leverages automatic differentiation through the unsteady numerical simulation procedure of a cardiovascular LPM to identify important parameters, relating to and including LV elastances for diseased and healthy left heart valves. The algorithm consists of a multi-compartment LPM of an entire cardiovascular system and 1st and 2nd order gradient-descent optimisation routines. To demonstrate the performance and accuracy of the proposed algorithm, synthetic data generated using the LPM is used as a proxy for clinical measurements. Only measurements which can be non-invasively acquired such as ventricular volumes, valvular flow rates and arterial pressures are utilised to discover the unknown system parameters. To identify a parameter subset to be optimised, a local sensitivity study is performed also using automatic differentiation to calculate the LV pressure and volume sensitivities relative to the cardiovascular model parameters. To lower costs of deploying the proposed algorithm and to enable reproducibility, the computer models are developed in free and opensource Julia 1.7.0 libraries namely \textit{DifferentialEquations.jl} \cite{Rackauckas2017}, \textit{ForwardDiff.jl} \cite{Revels2016} (automatic differentiation), \textit{Optim.jl} \cite{Morgensen2018} (optimisation framework) and \textit{Flux.jl} \cite{Innes2018} (first-order gradient-descent optimisers).

\section{Materials and methods}
\subsection{Overview of proposed model}

\begin{figure}[h!]
	\centering
	\includegraphics[width=\textwidth]{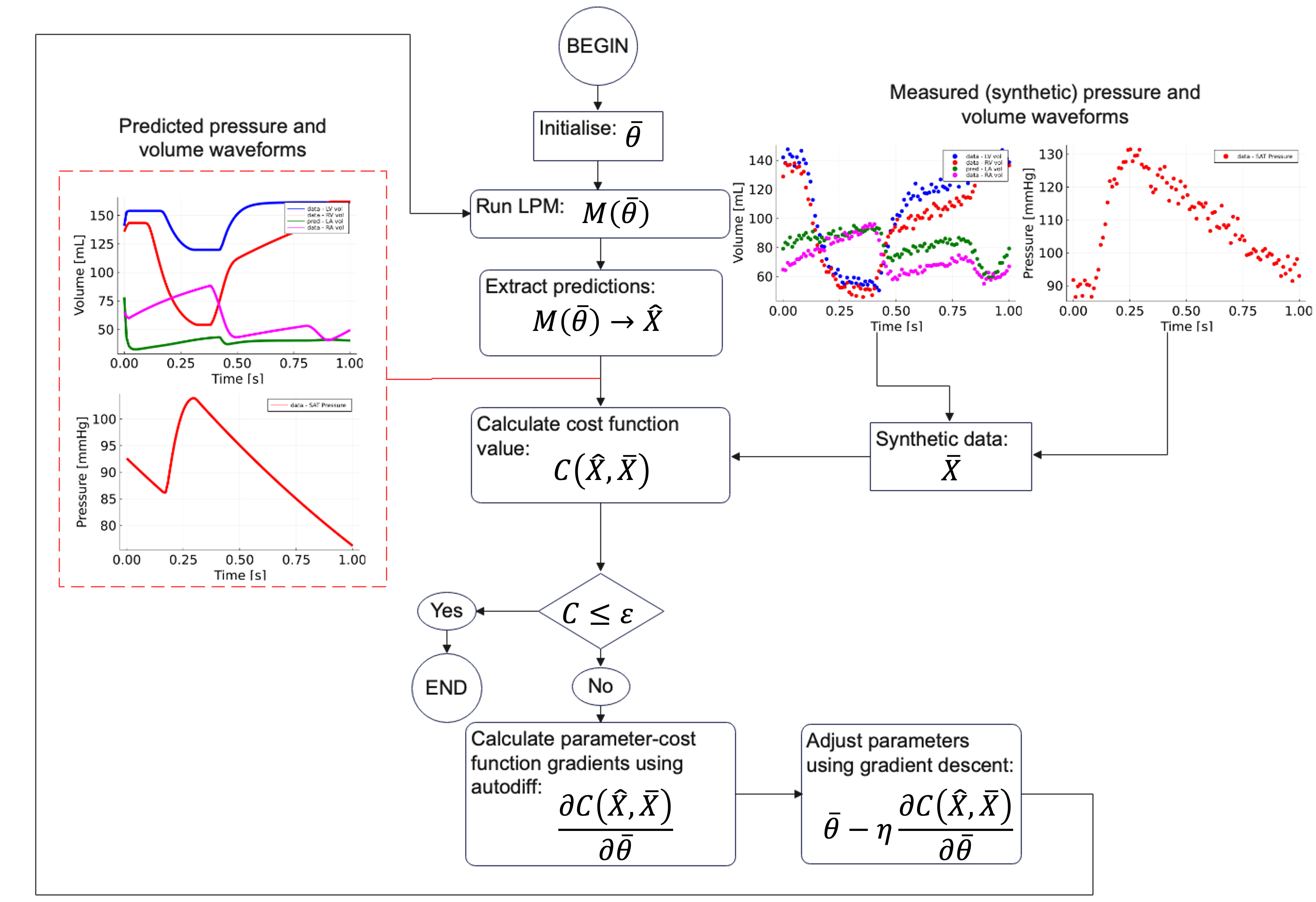}
	\caption{Computer model flow chart}
	\label{algorithm}
\end{figure}

In the present work, synthetic data of ventricular volume changes, valvular flow rates and arterial pressure waveforms for a single beat are generated using a multi-compartment LPM and serves as substitutions for actual clinical measurements. The benefit of using synthetic data is that the true parameters used to generate the measurements are known, thus the accuracy of the parameter optimisation model can be evaluated. The objective of the proposed model is to recover the LV elastances used during synthetic data generation, by utilising the substitute non-invasive measurements and the LPM equations. The development of the proposed model can be divided into two parts namely parameter subset selection and parameter optimisation (figure \ref{algorithm}).

Seeing as various cardiovascular parameters such as arterial resistances and unstressed LV volume influence the pressure-volume relationship of the LV and is typically unknown prior to invasive measurements, a sensitivity analysis using the LPM is performed. This is done to define a reduced parameter subset $\bar{\theta}$ which contains the LV elastances and important parameters which influence the LV pressure-volume loop. In the current work, the remainder of the model parameters which is not to be optimised are fixed to typical values found in literature. More on this later on in the paper. Once the parameter subset has been identified, it can be optimised by minimizing the error between the parameter-specific model predictions and the synthetic data.

In figure \ref{algorithm}, the program flow chart for the parameter optimisation is depicted. Initially $\bar{\theta}$ is randomly initialised to values within typical clinical ranges. Next, the parameters are used along with the LPM and an ODE solver to produce a solution for a single heartbeat period $\bar{X} = M(\bar{\theta})$. Using the parameter-specific mathematical model solution, the relevant waveforms $\tilde{X}$ is extracted from $\bar{X}$ and along with the synthetic data $\hat{X}$ fed into a cost function $C$. If the resulting cost function value is below the user specified convergence criteria $\epsilon$ the program exits. Conversely, if the cost function value is above $\epsilon$, the gradients of the cost function with respect to the model parameters $\frac{\partial C (\hat{X}, \bar{X})}{\partial \bar{\theta}}$ is calculated using forward-mode automatic differentiation. Next, $\frac{\partial C}{\partial \bar{\theta}}$ is used along with a gradient-descent optimiser to calculate the updated parameters $\bar{\theta}^{\text{new}}$ new which is then fed back to the LPM and ODE solver and the process is repeated. 

In the following subsections, the lumped parameter modelling, parameter optimisation and sensitivity analysis (reduced parameter subset selection) methods will be discussed in more detail.

\subsection{Mathematical modelling and synthetic datasets}
\subsubsection{Lumped parameter modelling}
The basis of the present work is a multi-compartment LPM of the entire cardiovascular system, which is depicted in figure \ref{lpm} below. The LPM developed in the present work consists of the four heart chambers and their respective downstream valves, the systemic loop and the pulmonary loop. 

\begin{figure}[h!]
	\centering
	\includegraphics[width=\textwidth]{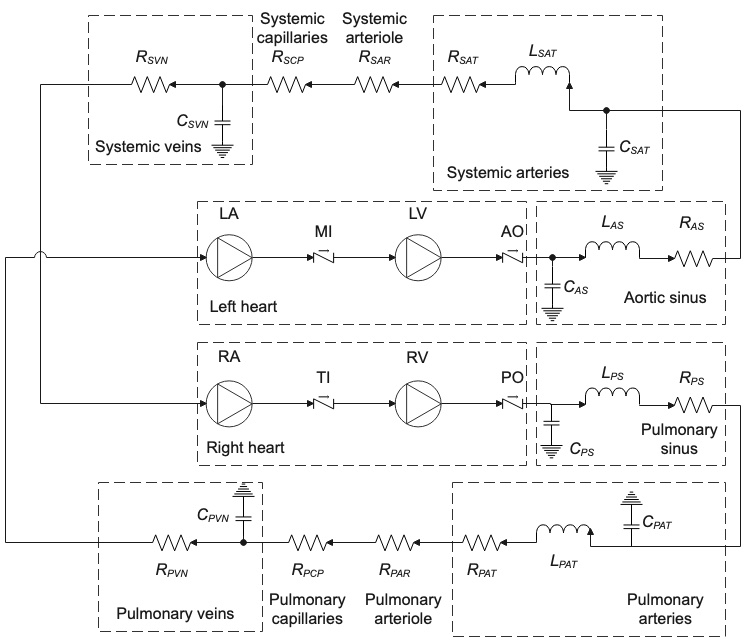}
	\caption{Cardiovascular multi-compartment lumped parameter model. Aortic – AO, mitral – MI, pulmonary – PO, tricuspid – TI, aortic sinus – AS, systemic arteries – SAT, systemic arterioles – SAR, systemic capillaries – SCP, systemic veins – SVN, pulmonary sinus – PS, pulmonary arteries – PAT, pulmonary arterioles - PAR, pulmonary capillaries – PCP, pulmonary veins - PVN}
	\label{lpm}
\end{figure}

The pressure-volume relationship of the heart chambers can be mathematically modelled using the Suga et al. \cite{Suga1973} relation as seen in equation \ref{heart_pressure}. In equation 1, $P_{lv} (t)$  [mmHg] is the LV pressure at time $t$, $P_{lv,0}$ is the unstressed LV pressure (set to a value of 1 for all heart chambers \cite{Korakianitis2006}), $e_{lv} (t)$  [s] is the LV time-varying elastance function, $V_{lv} (t)$ [mL] is the instantaneous LV volume and $V_{lv,0}$ is the unstressed LV volume. 

\begin{equation}
	P_{lv}(t) = P_{lv,0} + e_{lv}(t) \left( V_{lv}(t) - V_{lv,0} \right)
	\label{heart_pressure}
\end{equation}

The ventricle volume can be simulated using a simple mass balance relation over the ventricle control volume as shown in equation \ref{heart_vol}. In equation 2, $Q_{mi} (t)$ [mL/s] is the volume flow rate through the mitral valve and $Q_{ao} (t)$ is the flow rate through the aortic valve. 

\begin{equation}
	\frac{dV_{lv}}{dt} = Q_{mi}(t) - Q_{ao}(t)
	\label{heart_vol}
\end{equation}

The time-varying elastance function for the LV is calculated using the systolic and diastolic elastances $E_{lv,s}$, $E_{lv,d}$ [mmHg/mL] and the contractility activation function $f(t)$ as shown in equation \ref{heart_ventricle_elastance}.

\begin{equation}
	e_{lv}(t) = E_{lv,d} + \frac{E_{lv,s} - E_{lv,d}}{2}\cdot f(t)
	\label{heart_ventricle_elastance}
\end{equation}

In the present work the ventricle activation function used was taken from the work of Bozkurt \cite{Bozkurt2019} and is shown in equation \ref{heart_act_func_v}.

\begin{equation}
	f_i(t) = 
	\begin{cases}
		1 - \cos \left[ \left( \frac{t}{T_1} \pi \right) \right], & \text{if } 0 \leq t < T_1 \\
		1 + \cos \left[ \frac{t - T_1}{T_2 - T_1} \pi \right], & \text{if } T_1 \leq t < T_2 \\
		0, & \text{if } T_2 \leq t < T
		
	\end{cases}
	\label{heart_act_func_v}
\end{equation}

In equation \ref{heart_act_func_v}, $T_1=0.3 T$ [s] is the time at end of systole, $T_2=0.45 T$ is the time at end of ventricular relaxation \cite{Bozkurt2019} and $T$ is the heart beat period which in the present work was fixed to 1 [s]. These time intervals can typically be acquired from electrocardiogram (ECG) and Doppler echocardiography (DE) data. To model the right ventricle, equations \ref{heart_pressure}-\ref{heart_act_func_a} are similarly applied but with corresponding right heart parameters such as $E_{rv,s}$ and $E_{rv,d}$ and tricuspid and pulmonary valve flow rates $Q_{ti} (t)$ and $Q_{po} (t)$.

The atria pressures and instantaneous chamber volumes are calculated similarly to the ventricles (equations \ref{heart_pressure} and \ref{heart_vol}), but with different definitions for time-varying elastance functions and contractility activation functions. The time-varying left atrial (LA) elastance function is calculated using equation \ref{heart_atrium_elastance}. The right atrial elastance function is similarly calculated.

\begin{equation}
	e_{la}(t) = E_{la,min} + \frac{E_{la,max} - E_{la,min}}{2}\cdot f_a(t - D)
	\label{heart_atrium_elastance}
\end{equation}

In equation \ref{heart_atrium_elastance}, $E_{la,min}$ and $E_{la,max}$ are the minimal and maximal LA elastances, $f_a (t)$ is the atrial contractility activation function which in turn is calculated using equation \ref{heart_act_func_a} and $D = 0.04$ [s] is the time of atrial relaxation.

\begin{equation}
	f_a(t) =
	\begin{cases}
		0, & \text{if } 0 \leq t < T_a \\
		1 - \cos \left[ 2\pi \frac{t - T_a}{T - T_a} \right], & \text{if } T_a \leq t < T 
	\end{cases}
	\label{heart_act_func_a}
\end{equation}

In equation \ref{heart_act_func_a}, $T_a=0.8 T$ is the time at onset of atrial contraction. Similar to the ventricular time intervals, $T_a$ and $D$ can be measured using ECG and DE data. Table \ref{heart_table_a} below contains the nominal parameters \cite{Korakianitis2006a} used in the LPM.

\begin{table}[h!]
	\centering
	\caption{Heart model nominal parameters (values in parenthesis indicate upper and lower boundaries for sensitivity analysis and normalisation)}
	\begin{tabular} { c c c }
		\hline
		Parameters & Left heart & Right heart \\
		\hline
		&\emph{Atria} \\
		$E_{max}$ $\left[ \frac{mmHg}{mL} \right]$ & $0.25 (0.0,1.0)$ & $0.15 (0.0,1.0)$ \\
		$E_{min}$ $\left[ \frac{mmHg}{mL} \right]$ & $0.15 (0.0,0.5)$ & $0.15 (0.0,0.5)$ \\
		$V_{0}$ $\left[ mL \right]$ & $4 (1.0,20.0)$ & $4 (1.0,20.0)$ \\
		&\emph{Ventricles} \\
		$E_{d}$ $\left[ \frac{mmHg}{mL} \right]$ & $0.1(0.0,1.0)$ & $0.1(0.0,0.5)$ \\
		$E_{s}$ $\left[ \frac{mmHg}{mL} \right]$ & $2.5(0.5,5.0)$ & $1.15(0.5,5.0)$ \\
		$V_{0}$ $\left[ mL \right]$ & $5(1.0,20.0)$ & $10(1.0,50.0)$ \\
		\hline
	\end{tabular}	
	\label{heart_table_a}
\end{table}

The pressure-volume flow rate relation of the heart valve models was simulated using a orifice model proposed by Korakianitis and Shi \cite{Korakianitis2006}. For this heart valve model the flow rate through the $i^th$ valve is calculated using equation \ref{korak_q}, where $i = ao,mi,ti,po$ (aortic, mitral, tricuspid and pulmonary). 

\begin{equation}
	Q_i(t) = CQ \cdot A_{r}(t) \sqrt{\Delta P(t)}
	\label{korak_q}
\end{equation}

In equation \ref{korak_q}, $CQ$ is the flow coefficient which for the atrioventricular valves is set to 400 and for the semilunar valves to 350, $A_r (t)$ is the time-dependent area opening fraction of the valve, and $\Delta P (t)$ the pressure drop across the valve which in turn is calculated using equation \ref{korak_delta_p}.

\begin{equation}
	\Delta P(t) = 
	\begin{cases}
		P_{in}(t) - P_{ex}(t), \text{if } & P_{in} \geq P_{ex} \\
		P_{ex}(t) - P_{in}(t), \text{if } & P_{in} < P_{ex}
	\end{cases}
	\label{korak_delta_p}
\end{equation}

In equation \ref{korak_delta_p}, $P_{in} (t)$ is the instantaneous valve inlet pressure and $P_{ex} (t)$ the exit or downstream pressure. For example, the aortic valve inlet pressure would be the LV pressure $P_{lv} (t$ and the exit pressure the aortic sinus pressure $P_{AS} (t)$ as shown in figure \ref{lpm}. The valve area opening fraction is simulated using simple diode behaviour, where $A_r$ either takes a value of 0 or 1 depending on the pressure gradient across the specific heart valve as shown in equation \ref{korak_ar}.

\begin{equation}
	A_r(t) = 
	\begin{cases}
		1, \text{if } & P_{in}(t) \geq P_{ex}(t) \\
		0, \text{if } & P_{in}(t) < P_{ex}(t)
	\end{cases}
	\label{korak_ar}
\end{equation}  

The systemic and pulmonary loops each consists of five components which are sinuses, arteries, arterioles, capillaries and veins. For each of these compartments a set of equations are used to simulate their internal pressure and flow rate dynamics. Note that only the systemic loop equations will be discussed below, but the pulmonary loops solves similar equations with different parameters. The flow rate through the aortic sinus and its corresponding inlet pressure are simulated using two ordinary differential equations as shown as follows in equations \ref{AS_flow} and \ref{AS_P}.

\begin{equation}
	L_{AS} \frac{d Q_{AS}}{dt} = \left( P_{AS} - P_{SAT} \right) - R_{AS} Q_{AS}
	\label{AS_flow}
\end{equation}

\begin{equation}
	C_{AS} \frac{d P_{AS}}{dt} = Q_{AO} - Q_{AS}
	\label{AS_P}
\end{equation}

In equations \ref{AS_flow} and \ref{AS_P}, $L_{AS}$ $[\frac{mmHg \cdot s^2}{mL}]$ is the blood flow inertia through the aortic sinus, $Q_{AS}$ is the flow rate of blood through the sinus, $P_{AS}$ is the inlet sinus pressure, $P_{SAT}$ is the systemic arterial inlet pressure, $R_{AS}$ $[\frac{mmHg \cdot s}{mL}]$ is the sinus flow resistance and $C_{AS}$ $[\frac{mL}{mmHg}]$ is the sinus vessel wall compliance. The systemic arteries are simulated similarly to the aortic sinus. The equations governing the arterial pressure and flow rate are given in equations \ref{SAT_flow} and \ref{SAT_P}.

\begin{equation}
	L_{SAT} \frac{d Q_{SAT}}{dt} = \left( P_{SAT} - P_{SAR} \right) - R_{SAT} Q_{SAT}
	\label{SAT_flow}
\end{equation}

\begin{equation}
	C_{SAT} \frac{d P_{SAT}}{dt} = Q_{AS} - Q_{SAT}
	\label{SAT_P}
\end{equation}

The systemic arterioles and capillaries are simulated using only hydraulic resistance components due to the rigidity and small diameter of the vessel walls \cite{Tang2020}. The flow rates through these sections of the cardiovascular system are assumed to be steady, thus due to the mass conservation law the flow rates are all assumed to be equal $Q_{SAT}=Q_{SAR}=Q_{SCP}$. The pressure drop through these sections can, therefore, be calculated as:

\begin{equation}
	Q_{SAT} R_{SAR} = P_{SAR} - P_{SCP}
	\label{SAR_Q}
\end{equation}

\begin{equation}
	Q_{SAT} R_{SCP} = P_{SCP} - P_{SCN}
	\label{SCP_Q}
\end{equation}

The systemic veins are simulated using compliance-resistance components with negligible blood inertia \cite{Korakianitis2006}. The governing differential equation for the systemic vein inlet pressure can be seen in equation \ref{SVN_P}.

\begin{equation}
	C_{SVN} \frac{d P_{SVN}}{dt} = Q_{SCP} - Q_{SVN}
	\label{SVN_P}
\end{equation}

To calculate the flow rate through the venous system the hydraulic resistance expression is utilised. The driving pressure for the systemic vein flow rates is the difference in systemic vein inlet pressure and the right atrium pressure $P_{ra} (t)$, as seen in equation \ref{SVN_Q}.

\begin{equation}
    Q_{SVN} R_{SVN} = P_{SVN} - P_{ra}(t)
    \label{SVN_Q}
\end{equation}

The various model parameters used in the previous equations (\ref{AS_flow}-\ref{SVN_Q}) for both the systemic and pulmonary loops can be seen in table \ref{sys_pul_net_pars}.

\begin{table}[h!]
	\centering
	\caption{Systemic and pulmonary loop parameters (values in parenthesis indicate upper and lower boundaries for sensitivity analysis and normalisation) \cite{Korakianitis2006}}
	\begin{tabular} {c c c c}
	\hline
		Components & $R$ $\left[ \frac{mmHg \cdot s}{mL} \right]$ & $L$ $\left[ \frac{mmHg \cdot s^2}{mL} \right]$ & $C$ $\left[ \frac{mL}{mmHg} \right]$ \\
	\hline	
	\emph{Systemic network} \\
	AS & 0.003 ($3 \cdot 10^{-4}$, 0.03) & $6.2\cdot 10^{-5}$ ($1 \cdot 10^{-4}$, 0.001) & 0.08 (0.008,0.8) \\
	SAT & 0.05 (0.005,1.0) & 0.0017 ($1.7 \cdot 10^{-3}$, 0.017) & 1.6 (0.16,3.2)  \\
	SAR & 0.5 (0.05, 1.0) & - & -  \\
	SCP & 0.52 (0.05, 1.0) & - & -  \\
	SVN & 0.075 (0.0075, 0.75) & - & 20.5 (5.0, 50.0)  \\
	\emph{Pulmonary network} \\
	PS & 0.002 (0.0002, 0.002) & $5.2 \cdot 10^{-5}$ ($1\cdot10^{-4}$, $1\cdot10^{-3}$) & 0.18 (0.018, 2.0)  \\
	PAT & 0.01 (0.001, 0.1) & 0.0017 ($1.7 \cdot 10^{-3}$, 0.017) & 3.8 (0.038, 6.0)  \\
	PAR & 0.05 (0.005, 1.0) & - & -  \\
	PCP & 0.05 (0.025, 2.0) & - & - \\
	PVN & 0.006 ($6 \cdot 10^{-4}$, 0.01) & - & 20.5 (5.0, 50.0)  \\
	\hline
	\end{tabular} \\
	\label{sys_pul_net_pars}
\end{table}

To simulate the dynamics of the cardiovascular system, the differential-algebraic equations (DAEs) for the time-varying elastance model (equations \ref{heart_pressure}-\ref{heart_act_func_a}), the heart valve model equations (equations \ref{korak_q}-\ref{korak_ar}) and vasculature equations (equations \ref{AS_flow}-\ref{SVN_Q}) for the left and right heart chambers, valves and systemic and pulmonary loops are solved simultaneously using an ODE integrator. For the current work the Tsitouras 5/4 Runge-Kutta variable time-step method \cite{Tsitouras2011} is applied with relative and absolute tolerances set to 1E-4 and 1E-6 respectively. In total the model solves for 14 dependent variables and uses the parameters from tables \ref{heart_table_a} and \ref{sys_pul_net_pars} as model inputs along with the initial conditions vector $\bar{X}_{init}$. These 14 dependent variables are defined by the model equations for the four heart chambers, four heart valves and systemic and pulmonary loops as discussed in section 2.2.1. For a more detailed discussion of the model equations the interested reader is referred to \cite{Laubscher2022}. Once the model is solved, the corresponding output has the size $M(\tilde{\theta}, \bar{X}_init) = \tilde{X} \in \mathcal{R}^{14 \times N_t}$ where $\tilde{\theta}$ is a vector containing all the parameters mentioned above, $\tilde{X}$ is the model solution matrix and $N_t$ is the total number of time steps taken by the ODE integrator over the simulation time $T = 1$ [s]. Note, that the parameters to be optimised is only a subset of the complete parameter vector thus $\bar{\theta} \in \bar{\theta}$ and similarly $\bar{X}$ is only a subset of the solution matrix thus $\bar{X} \in \tilde{X}$.

\subsubsection{Synthetic data and lumped parameter model initial conditions}
Clinical measurement data is typically discrete values sampled at a certain frequency, therefore, the synthetic data matrix has a size of $\hat{X} \in \mathcal{R}^{d \times N_t^*}$, where $d$ is the number of different types of measurements and $N_t^*$ the number of sampled time steps for each performed measurement. In the present work $N_t^*=100$ which equivalent to a sampling frequency of 100 Hz. To compare the simulated trends to the synthetic data, a subset of the simulated values is sampled from $\tilde{X}$ at time steps corresponding to the synthetically measured time stamps, thus resulting in the simulation subset $\bar{X} \in \mathcal{R}^{d \times N_t^*}$.

In the current work, three measurement datasets containing different combinations of typical clinical non-invasive measurements are evaluated to find the best-performing configuration that results in the smallest error between the actual model parameters and optimised ones. Table \ref{datasets} shows the dimensions and measurement collections of these three $\hat{X}$ datasets.

\begin{table}[h!]
	\centering
	\caption{Time-varying elastance model parameters for atriums. Ref. \cite{Bozkurt2019}, \cite{Korakianitis2006a}}
	\begin{tabular} { c c c }
		\hline
		Description & Dimensions & Simulated measurements \\
		\hline
        Dataset 1 (D1) & $d=5$, $N_t^*=100$ & $P_{SAT}$, $V_{la}$, $V_{lv}$, $V_{ra}$, $V_{rv}$ \\
        Dataset 2 (D2) & $d=5$, $N_t^*=100$ & $P_{SAT}$, $Q_{ao}$, $Q_{mi}$, $Q_{po}$, $Q_{ti}$ \\
        Dataset 3 (D3) & $d=9$, $N_t^*=100$ & D1 + D2 \\
		\hline
	\end{tabular}	
	\label{datasets}
\end{table}

To measure the clinical trends in table 3, different medical equipment can be used. For the current study the following measurements are proposed for further clinical studies. For the heart chamber volume trends, 3D echography or magnetic resonance imaging (MRI) could be utilised and Doppler echography to measure the heart valve flow rates. The brachial arterial pressure can be continuously measured using a CNAP Monitor \cite{Zhang2020a} and the volume-clamp method.

To simulate the dynamics of the LPM dependent variables such as LV volume changes, systemic arterial pressures and venous pressures, using the model equations, 14 initial conditions are required. The initial conditions vector can be defined as $\bar{X}_{init} = [ V_{lv}^{init}, V_{la}^{init}, P_{AS}^{init},Q_{AS}^{init}, P_{SAT}^{init},$ 

$Q_{SAT}^{init}, P_{SVN}^{init}, V_{rv}^{init}, V_{ra}^{init}, P_{PS}^{init}, Q_{PS}^{init}, P_{PAT}^{init}, Q_{PAT}^{init}, P_{PVN}^{init} ]$. In a clinical application of the current work, these initial conditions should also be extracted from non-invasive measurements. The initial heart chamber volumes, $V_{lv}^{init}, V_{la}^{init},V_{rv}^{init}$ and $V_{ra}^{init}$, can be directly taken as the first entries in the corresponding synthetic data trends. The initial aortic  and pulmonary sinus flow rates, $Q_{AS}^{init}$ and $Q_{PS}^{init}$ , are assumed to be equal to the initial entries in the synthetic datasets for the aortic and pulmonary valvular flow rates. The initial systemic arterial pressure $P_{SAT}^{init}$ can be extracted from the synthetic dataset trends. The initial aortic sinus pressure $P_{AS}^{init}$ is assumed to be equal to the initial systemic arterial pressure. For the remaining initial conditions, additional clinical measurements are required. The total systemic and pulmonary arterial flow rates $Q_{SAT}^{init}$ and $Q_{PAT}^{init}$ is estimated using the LV and RV stroke volume measured using Doppler echography and divided by the heartbeat interval time. The initial pulmonary and systemic venous pressures $P_{SVN}^{init}$ and $P_{PVN}^{init}$ are important initial conditions, seeing as it drives the blood flow into the atria. These pressures are difficult to estimate non-invasively, therefore, in the present work the initial venous pressures are optimised along with the other model parameters. Similarly, the pulmonary sinus and arterial pressure initial conditions $P_{PS}^{init}$ and $P_{PAT}^{init}$ are also difficult to estimate non-invasively, and therefore, are also optimised along with the model parameters.

\subsection{Parameter optimisation}
The parameter vector containing the unknown cardiovascular system quantities is in essence a row vector of values which individually each have a significant influence on the LV pressure-volume relationship. How the parameters in this vector are selected will be discussed in section 2.4. An example of the parameter vector is shown in equation \ref{eq18}.

\begin{equation}
 \bar{\theta} = [E_{lv,s},E_{lv,d},…,R_{AS},…,C_{PVN} ] 
 \label{eq18}
\end{equation}

To find $\bar{\theta}$ which minimizes the differences between the simulation results and measurement data, a sum squared error (SSE) cost function is defined for each of the measurements, as shown in equation \ref{eq19} for the $j=1,2,..,d$ measurement types (e.g. LV volume, arterial pressure or aortic valve flow rate).

\begin{equation}
    J(\bar{x}_j, \hat{x}_j) = \left( \sum_{i=1}^{N_t^*} (\bar{x}_j^i(\bar{\theta}) - \hat{x}_j^i)^2 \right)_j
    \label{eq19}
\end{equation}

In equation \ref{eq19}, $\bar{x_j^i}$  is the $j^{th}$ simulation output at time step $i$ and $\hat{x}_j^i$ is the $j^{th}$ synthetically measured value (e.g. arterial pressure or LV volume) at time step $i$. The objective of the optimisation routine is to minimise the sum of the individual measurement type cost functions as shown in equation \ref{eq20}.

\begin{equation}
    C(\bar{X}(\bar{\theta}), \hat{X}) = \sum_{i=1}^{d} \left( J(\bar{x}_i (\bar{\theta}), \hat{x}_i) \right)
    \label{eq20}
\end{equation}

To speed up model convergence, the parameter vector entries are normalised using min-max scaling and user specified upper and lower parameter boundaries (see tables 1 and 2). Min-max scaling results in parameters having numeric values between 0-1. The scaling transformation can be seen in equation \ref{eq21}, where $\bar{\theta}^*$ is the scaled parameter vector, $\bar{\theta}_{lb}$ is a vector of lower boundary parameter values and $\bar{\theta}_{ub}$ is a vector of upper boundary parameter values. Note that during the optimisation routine, the parameter vector is rescaled before being sent to the LPM and ODE integrator routine.

\begin{equation}
    \bar{\theta}^* = \frac{\bar{\theta} - \bar{\theta}_{lb}}{\bar{\theta}_{ub} - \bar{\theta}_{lb}}
    \label{eq21}
\end{equation}

Due to the difference in magnitudes of volumes, flow rates and pressures, the simulation outputs and measurement data are also scaled to values between 0-1 before being used in the cost function calculations. Min-max scaling is again used along with the maximum and minimum measured values for each measurement type. For example, the scaled simulated arterial pressures can be calculated using equation \ref{eq22}, where the subscript $p$ indicates arterial pressure values being compared from the simulation output and synthetic measurement vectors $\bar{x}$ and $\hat{x}$ respectively.

\begin{equation}
    \bar{x}_p^* = \frac{\bar{x}_p - \text{min}(\bar{x_p})}{\text{max}(\bar{x_p}) - \text{min}(\bar{x_p})}
    \label{eq22}
\end{equation}

To find the parameters which minimises the cost function $C(\bar{\theta})$, a combination of first- and second-order gradient descent optimisers are utilised. The adaptive moment estimation (Adam) first-order optimiser \cite{Kingma2015} was selected. The Adam optimiser algorithm is shown in equation \ref{eq23}.

\begin{equation}
    \begin{gathered}
        \bar{m} \leftarrow \beta_1 \bar{m} + (1 - \beta_1) \nabla_\theta C(\bar{\theta}) \\
        \bar{s} \leftarrow \beta_1 \bar{s} + (1 - \beta_2) \nabla_\theta C(\bar{\theta}) \otimes \nabla_\theta C(\bar{\theta}) \\
        \bar{m} \leftarrow \frac{\bar{m}}{1 - \beta_1^t} \\
        \bar{s} \leftarrow \frac{\bar{s}}{1 - \beta_2^t} \\
        \bar{\theta}^{\text{new}} \leftarrow \bar{\theta} - \eta \bar{m} \otimes \sqrt{(\bar{s} + \epsilon)^{-1}}
        \label{eq23}
    \end{gathered}
\end{equation}

The scaling $\bar{s}$ and momentum $\bar{m}$ matrices are initialized to 0 at the start of the Adam training algorithm, $t$ is the iteration counter, $\epsilon = 10^{-8}$ is the smoothing term, $\beta_1$ is the momentum decay hyperparameter and is set to 0.9 and $\beta_2$ is the scaling hyperparameter and is set to 0.999. In equation \ref{eq23}, $\nabla_\theta C(\bar{\theta})$ are the gradients of the cost function with respect to the optimisation parameters.  For the optimisation runs, the learning rate parameter $\eta$ is fixed to a value of 0.005. In the development of the optimisation model, other optimisers were tested such as the standard gradient descent, momentum and Nestorov algorithms, but to stabilise these simulations, a small learning rate of 1E-5 had to be used. Therefore, it was decided to only use the Adam optimiser due to its superior stability for the current problem.

As mentioned, in addition to using first-order optimisation, a second-order optimiser was also applied to find the optimal parameters. Second-order optimisers use the Hessian in addition to the gradients to determine the parameter step direction and step lengths during optimisation \cite{Kochenderfer2019a}. In the current work, the quasi-Newton Limited-memory Broyden-Fletcher-Goldfarb-Shanno (L-BFGS) \cite{Nocedal1980} optimiser was utilised, which does not directly calculate the Hessian but rather approximates it using the first-order gradients.

For both the above mentioned first- and second-order optimisers, the cost function gradients with respect to the parameters $\nabla_\theta C(\bar{\theta})$, are required. Forward-mode automatic differentiation is applied to calculate these gradients directly by tracking both values and derivatives as they are passed through the computational graphs created by the various mathematical operations. The \textit{ForwardDiff.jl} library used in the present work, is capable of traversing any native Julia code, therefore, it is capable of differentiating through the ODE integrator to calculate the required gradients in a computationally efficient manner, as will be demonstrated in the results section. 

\subsection{Sensitivity analysis}
To identify the model parameters (tables 1 and 2) which have a significant influence on the LV behaviour, a local sensitivity analysis is performed, excluding model initial conditions. This entails calculating the relative sensitivity percentages for each parameter using the LPM, the nominal values, upper and lower parameter boundary values and forward-mode automatic differentiation through the LPM solution. 

The main outputs which underpin the LV contractility is its time-dependent volume and internal pressure changes. Therefore, to calculate the parameter sensitivity of these values, the gradients of the mean pressure and volume with respect to each parameter, listed in tables 1 and 2, are calculated centred around the nominal values. Seeing that the model parameters vary significantly in order of magnitude and units, the mean LV pressure and volume parameter gradients were multiplied by their respective parameter ranges as seen below. Equations \ref{eq24} and \ref{eq25}, shows the mean LV pressure and volume sensitivity indices calculated for the $i^{th}$ parameter $\theta_i$.

\begin{equation}
    SI_{P,i} = \left|\frac{\partial \left( \frac{1}{N_t^*} \sum_{j=1}^{N_t^*} P_{lv}^j \right)}{\partial \theta_i}\right| \cdot \left( \theta_{i,ub} - \theta_{i,lb} \right)
    \label{eq24}
\end{equation}

\begin{equation}
    SI_{V,i} = \left|\frac{\partial \left( \frac{1}{N_t^*} \sum_{j=1}^{N_t^*} V_{lv}^j \right)}{\partial \theta_i}\right| \cdot \left( \theta_{i,ub} - \theta_{i,lb} \right)
    \label{eq25}
\end{equation}

In equations \ref{eq24} and \ref{eq25}, $P_{lv}^j$ is the LPM simulated LV pressure at time sample $j$, similarly $V_{lv}^j$ is the simulated LV volume, $\theta_{i,ub}$ is the upper parameter boundary value for parameter $i$ and similarly $\theta_{i,lb}$ the lower parameter boundary value. The relative sensitivity percentages are then calculated using equations \ref{eq26} and \ref{eq27}, where $n_{tot}$ is the total number of parameters listed in tables 1 and 2.

\begin{equation}
    SP_{P,i} = 100\% \cdot \frac{SI_{P,i}}{\sum_{j=1}^{n_{tot}} SI_{P,j}}
    \label{eq26}
\end{equation}

\begin{equation}
    SP_{V,i} = 100\% \cdot \frac{SI_{V,i}}{\sum_{j=1}^{n_{tot}} SI_{V,j}}
    \label{eq27}
\end{equation}

The parameters which describes 99\% of the variations in LV volume and pressure were used in the optimisation algorithm, as will be discussed in section 3.1. 

\subsection{Case studies}
To showcase the capabilities of the proposed optimisation model, three case studies are investigated. The synthetic data generated for case study 1, is for healthy heart valves using the nominal parameters listed in tables 1 and 2. To recreate clinical data, normally distributed noise \cite{Bjordalsbakke2022} was added to the synthetically generated arterial pressure, chamber volumes and heart valve flow data which forms the set of measurements used to optimise the model parameters. The standard deviation used for noise generation was set to 3\% of the mean value for the volume and flow rate measurements and 1\% for the arterial pressure measurement. Case study 2 used the nominal parameters for data generation but with the aortic valve maximum opening fraction set to 0.222 which corresponds to a valvular flow area of 1 [$cm^2$]. Case study 3, is for a regurgitant mitral valve. For this case, the minimum flow area opening fraction of the mitral valve was set to 0.05. For each of the case studies, the three datasets shown in table 3 was used in the optimisation model to gauge which resulted in the best parameter estimation accuracy. Figure \ref{cases} below shows the synthetically generated chamber volumes, arterial pressures and valve flow rates generated for the three case studies.

\begin{figure}[h!]
	\centering
	\includegraphics[width=\textwidth]{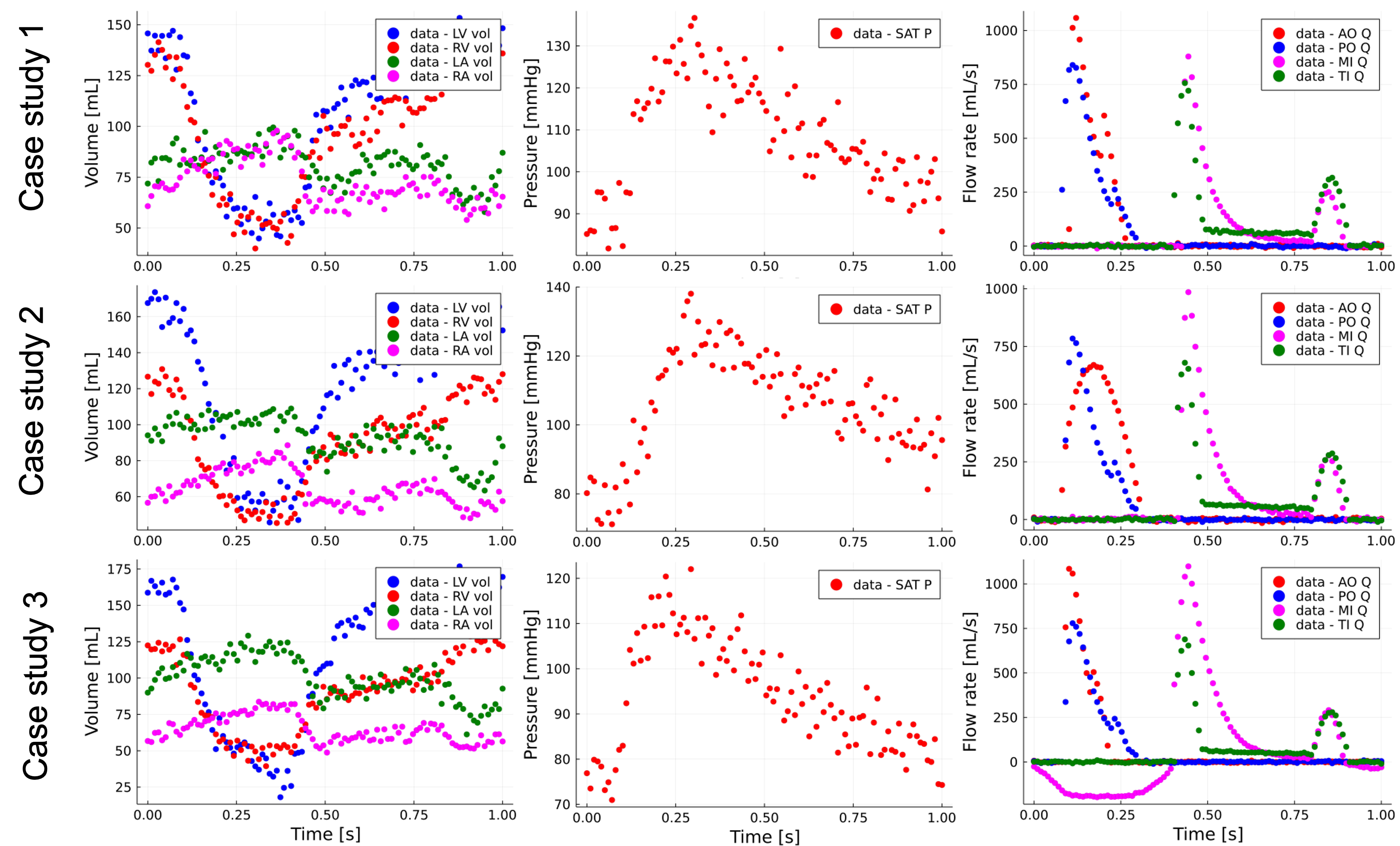}
	\caption{Case studies synthetic data}
	\label{cases}
\end{figure}

\section{Results and discussion}
\subsection{Sensitivity analysis}
\begin{figure}[h!]
	\centering
	\includegraphics[width=\textwidth]{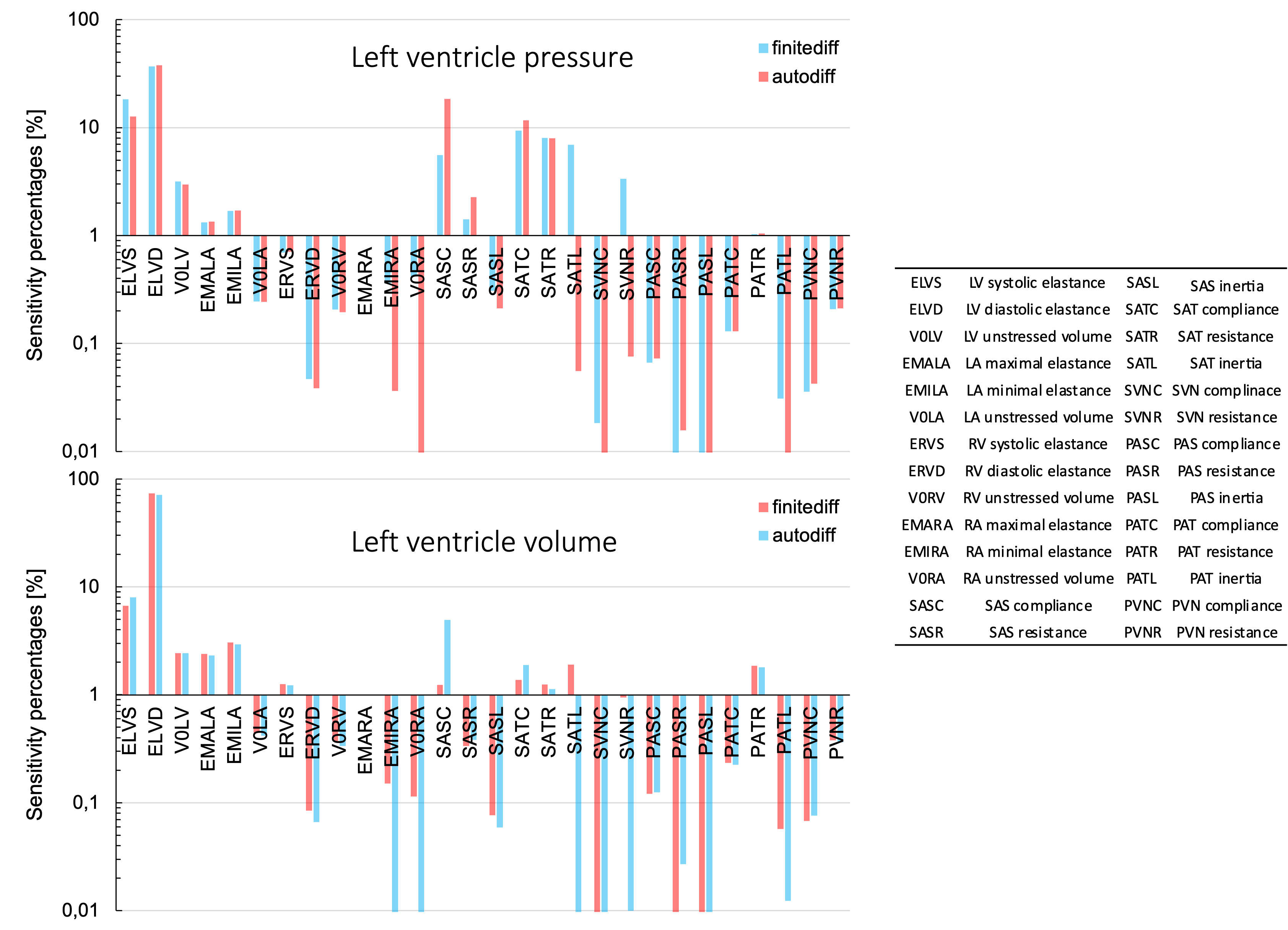}
	\caption{Sensitivity analysis results using autodiff and finite differences}
	\label{sens_res}
\end{figure}

Figure \ref{sens_res} shows the results of the sensitivity analysis performed using both automatic differentiation and finite differences to calculate the relevant gradients for case study 1 (healthy heart). From the autodiff sensitivity analysis results, it is seen that the LV diastolic elastance parameter has the largest effect on both LV mean pressure (37.8\%) and volume (71.2\%). The parameters with the second and third largest effects on the mean LV pressure is aortic sinus compliance (18.5\%) and LV systolic elastance (12.7\%). For the mean LV volume the second and third rated parameters are LV systolic elastance (8\%) and aortic sinus compliance (5\%). Using the sensitivity analysis results, 12 parameters were identified that explained 99\% of the variation in mean LV pressure and volume. The selected 12 optimisation parameters contained in $\bar{\theta}$ along with the initial systemic and pulmonary venous pressures and the pulmonary sinus and arterial pressures, can be seen in equation \ref{eq28}.

\begin{equation}
\begin{gathered}
    \bar{\theta} = [E_{lv,s}, E_{lv,d}, E_{rv,s}, E_{rv,d}, E_{la,max}, E_{la,min}, V_{lv,0}, C_{AS}, \\ R_{AS}, C_{SAT}, R_{SAT}, R_{PAT}, P_{PVN}^{init}, P_{SVN}^{init}, P_{PS}^{init}, P_{PAT}^{init}]
\end{gathered}
    \label{eq28}
\end{equation}

The above sensitivity analysis was also performed for case studies 2 and 3, and it was found that the algorithm identified the same 12 parameters, albeit, with a slightly different distribution of sensitivity percentages. 

Furthermore, the results in figure \ref{sens_res}, show that the parameter sensitivity percentages calculated for the LV, LA and RV using finite differences are similar to the values calculated using automatic-differentiation. Conversely, the finite differences method over- and under-predicts the importance of the RA parameters and various pulmonary and systemic loop parameters, highlighting the errors that could be made by approximating gradients using the finite differences approach.

\subsection{Case studies}
For each of the three case studies discussed in section 2.5, the unknown model parameters $\bar{\theta}$ are optimised using three optimisation approaches. The first approach, applies Adam optimisation and forward-mode automatic differentiation for 1000 training iterations. The second approach uses a combination of the Adam and LBFGS optimisers, a standard technique used in scientific machine learning \cite{Laubscher2021d}, \cite{Markidis2021}.  The model is first trained for 50 iterations using the Adam optimiser after which the second-order optimiser is utilised. The third approach simply applies the Adam optimiser for 1000 iterations using central finite differences to approximate the required gradients. For each of the case studies and optimisation approaches the effect of synthetic dataset contents (available measurements) was also investigated, by optimising the model parameters using the datasets listed in table 3. For each of the optimisation runs the initial model parameter vector was kept constant so as to compare the performance of the different optimisation permutations.

\begin{equation}
    \bar{\theta}_{init} = [1,0.01,1,0.01,0.5,0.49,0.1,0.01,1,1,0.1,10,10,25,25]
    \label{eq29}
\end{equation}

\subsubsection{Case study 1}
\begin{figure}[h!]
	\centering
	\includegraphics[width=\textwidth]{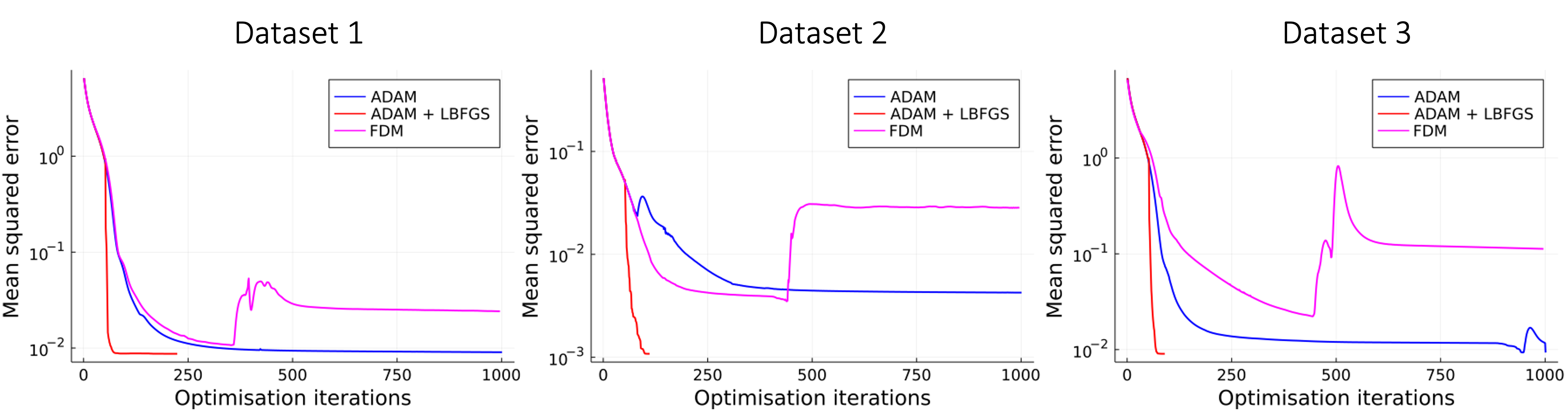}
	\caption{Mean squared error optimisation history for case study 1 and all datasets}
	\label{case1_hist}
\end{figure}

Figure \ref{case1_hist}, shows the mean squared error (MSE) training history for the three datasets and optimisation approaches using case study 1 model settings and measurement data. The results indicate that the finite difference optimisation method (FDM) yields the largest errors compared to the other optimisation approaches. Note FDM also uses the Adam optimiser but the gradients are approximated using finite differences. Comparing the training history of the Adam and hybrid Adam-LBFGS optimisation models (both using forward-mode automatic differentiation), shows that the latter has superior performance, in that it results in lower overall MSEs (especially for dataset 2) and converges in a fewer number of iterations. For the Adam-LBFGS approach, the rapid reduction of the MSE after the initial 50 iterations can clearly be observed. It should be noted that for datasets 1 and 3, the Adam optimiser approach eventually yields MSE values close to the Adam-LBFGS values. 

To compare the computational efficiency of the different approaches the total run times are recorded. The optimisations were performed using a 2020 Apple MacBook Pro M1. The FDM model calculated 1000 iterations in 230 seconds, the Adam approach completed the same number of iterations in 71 seconds and the Adam-LBFGS approach completed its ± 200 iterations in approximately 40 seconds. 

To compare the accuracy of the different permutations (optimisation approaches and datasets), the absolute percentage errors (APEs) between the true nominal parameters and the optimised values are calculated. Equation \ref{eq30} shows the formula used to calculate the APE for the $i^{th}$ parameter, where $\theta_i$ is the optimised parameter and $\theta_{i,true}$ the true nominal parameter. 

\begin{equation}
    APE = 100\% \cdot \frac{\left| \theta_i - \theta_{i,true} \right|}{\theta_{i,true}}
    \label{eq30}
\end{equation}

The absolute and mean absolute percentage error (MAPE) results for case study 1 with the different optimisation approaches and datasets are shown in table \ref{case1_table}. Note the shorthand for datasets, D1, D2 and D3 corresponds to datasets 1,2 and 3 respectively. In the tabular results, it is observed that both the Adam and Adam-LBFGS approaches are capable of estimating the heart parameters (first seven row entries) with  APEs between 0-10\%, excluding the unstressed LV volume, whereas the FDM approach struggles to accurately estimate these values.

\begin{table}[htbp]
\centering
  \caption{Absolute percentage errors per parameter for case study 1 and different synthetic datasets}
    \begin{tabular}{p{1cm} p{1cm} p{1cm} p{1cm} p{1cm} p{1cm} p{1cm} p{1cm} p{1cm} p{1cm}}
    \hline
    Par. & D1+ ADAM & D1+ LBFGS & D1+ FDM & D2+ ADAM & D2+ LBFGS & D2+ FDM & D3+ ADAM & D3+ LBFGS & D3+ FDM \\
    \hline
         $E_{lv,s}$ & 3.01  & 0.83  & 2.84  & 9.79  & 0.06  & 7.46  & 6.37  & 0.23  & 2.34 \\
         $E_{lv,d}$ & 7.11  & 4.67  & 12.60 & 8.25  & 3.05  & 1.37  & 4.22  & 5.17  & 4.26 \\
          $E_{rv,s}$ & 6.68  & 5.71  & 15.24 & 1.06  & 1.79  & 3.42  & 0.89  & 2.54  & 31.66 \\
          $E_{rv,d}$ & 0.48  & 0.52  & 0.39  & 0.32  & 8.37  & 12.04 & 0.29  & 0.77  & 2.02 \\
          $E_{la,max}$ & 8.01  & 5.10  & 9.39  & 8.58  & 5.05  & 19.29 & 5.86  & 6.03  & 5.51 \\
         $E_{la,min}$ & 6.65  & 7.29  & 8.13  & 6.30  & 3.41  & 15.11 & 4.70  & 7.56  & 7.68 \\
          $V_{lv,0}$ & 18.78 & 33.74 & 61.13 & 5.89  & 39.26 & 23.83 & 35.22 & 38.41 & 63.23 \\
         $C_{AS}$ & 54.17 & 27.45 & 81.37 & 178.93 & 2.90  & 71.83 & 1.66  & 4.55  & 87.34 \\
         $R_{AS}$ & 77.08 & 42.60 & 188.27 & 113.63 & 3.89  & 57.03 & 96.30 & 6.73  & 44.96 \\
          $C_{SAT}$ & 8.59  & 7.91  & 15.60 & 21.02 & 7.16  & 16.69 & 8.93  & 7.54  & 7.95 \\
          $R_{SAT}$ & 7.74  & 8.49  & 13.17 & 15.03 & 10.95 & 15.43 & 8.34  & 9.51  & 3.90 \\
          $R_{PAT}$ & 24.09 & 1.70  & 29.46 & 6.79  & 4.17  & 14.28 & 12.53 & 3.32  & 33.36 \\
          $P_{PVN}^{init}$ & 2.81  & 2.78  & 2.99  & 2.41  & 5.80  & 5.80  & 2.95  & 2.86  & 5.73 \\
          $P_{SVN}^{init}$ & 3.62  & 9.28  & 5.50  & 4.50  & 2.86  & 0.64  & 2.03  & 2.81  & 5.84 \\
          $P_{PS}^{init}$ & 9.12  & 7.83  & 21.93 & 0.50  & 4.97  & 7.10  & 1.69  & 2.03  & 30.84 \\
          \hline
          MAPE & 15.86 & 11.06 & 31.20 & 25.53 & 6.91  & 18.09 & 12.80 & 6.67  & 22.44 \\
    \hline
    \end{tabular}
  \label{case1_table}
\end{table}

The two configurations with the lowest overall MAPEs are D2 and D3 optimised using the Adam-LBFGS approach. For both these configurations the $V_{lv,0}$ parameter have error percentages above 30\%, but studying the sensitivity analysis results, it can be seen that the unstressed LV volume has a relative small effect on the ventricular performance for the analysed range. The result of this insensitivity to the LV unstressed volume can be seen in figure \ref{case1_plots} below, where it is shown that the Adam-LBFGS model configurations using D2 and D3 can accurately infer the observed arterial pressure and heart chamber volume waveforms and the unobserved LV pressure waveform by solving the LPM equations with the optimised parameters. Furthermore, it is seen from table 4, that although the unstressed volume error is large, the remaining parameter estimate errors are relatively low. For sake of brevity the other results for case study 1 are not provided.

\begin{figure}[h!]
	\centering
	\includegraphics[width=\textwidth]{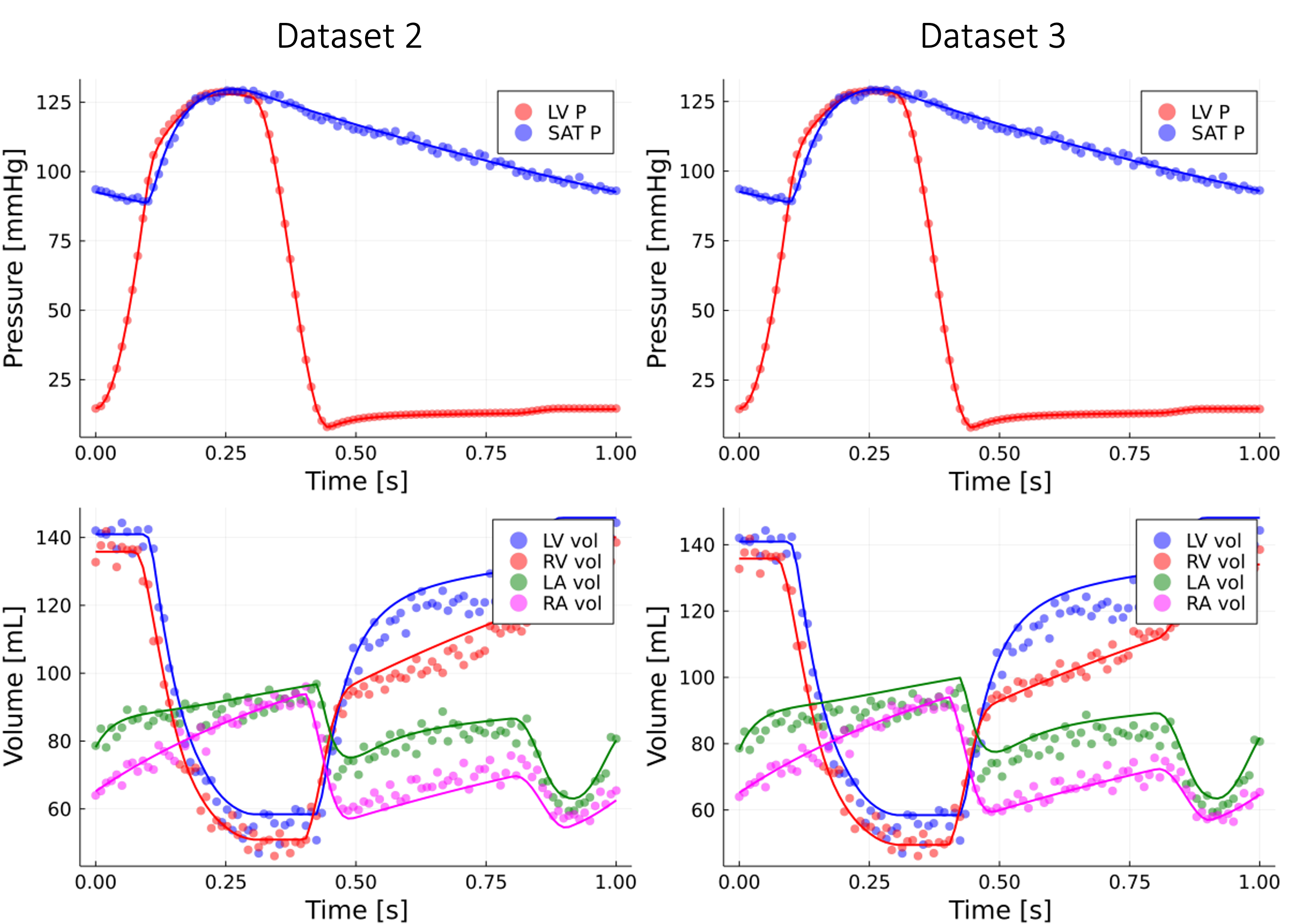}
	\caption{Captured heart chamber volumes, arterial pressure and inferred LV pressure using Adam and Adam-LBFGS models for case study 1 using datasets 2 and 3 (markers – synthetic measurement data, line – simulated using optimised parameters)}
	\label{case1_plots}
\end{figure}

\subsubsection{Case studies 2 and 3}
\begin{figure}[h!]
	\centering
	\includegraphics[width=\textwidth]{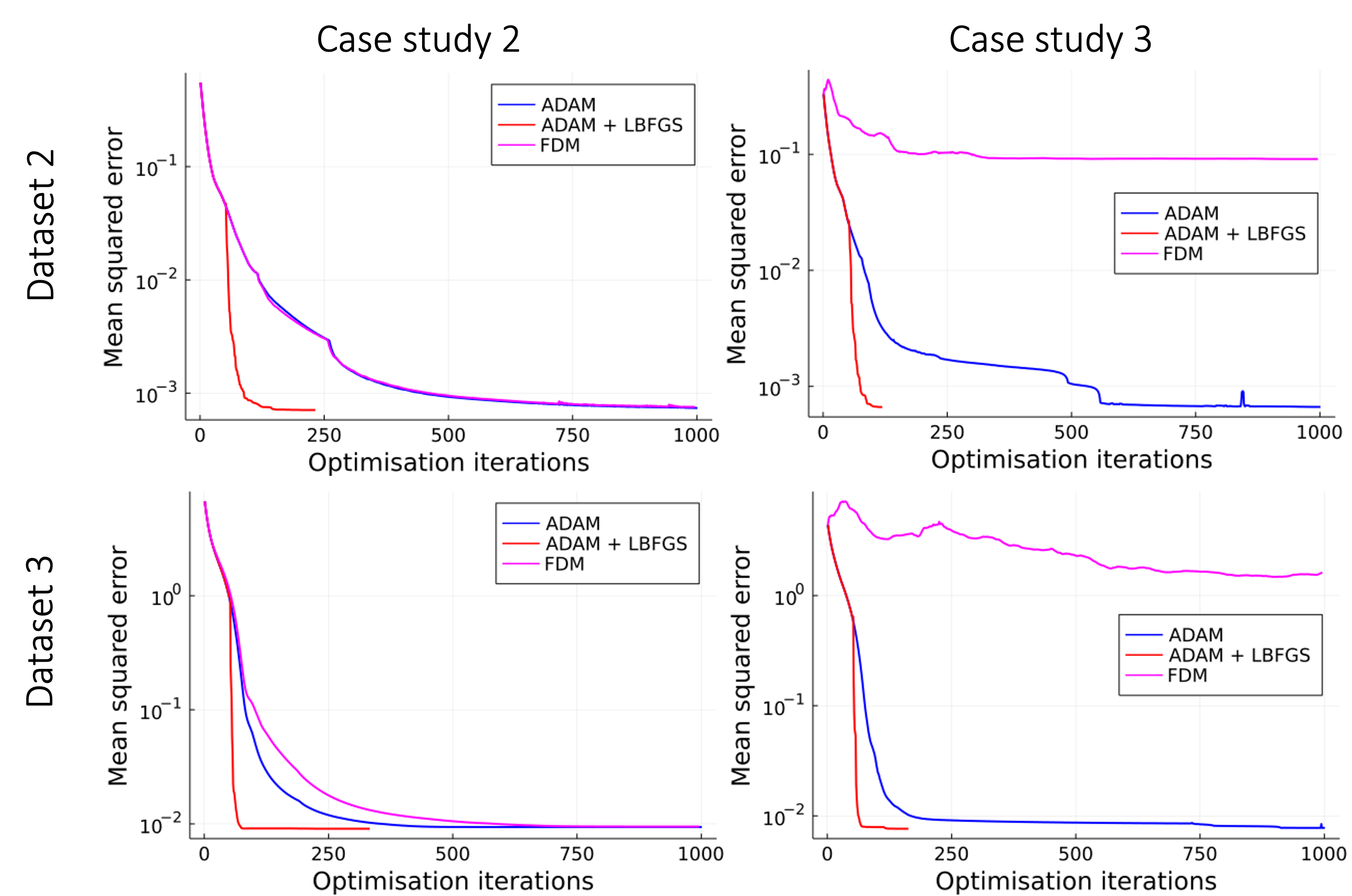}
	\caption{Mean squared optimisation error history for case studies 2 and 3 using datasets 2 and 3}
	\label{case23_hists}
\end{figure}

Seeing as the results generated using D2 and D3 clearly outperform the D1 results for case study 1, the remainder of this section will only use D2 and D3 for parameter estimations. 
Figure \ref{case23_hists} shows the training history for case studies 2 (aortic stenosis) and 3 (mitral regurgitation) using D2 and D3 along with the mentioned optimisation approaches. The results again show that the Adam-LBFGS approach outperforms the other two optimisers in terms of converged MSE values. The converged MSE values generated using D2 for both case studies are approximately $10^{-3}$, whereas, the MSE values achieved using D3 for model training are approximately $10^{-2}$. This leads to the conclusion that the addition of the heart chamber volume waveforms in the measurement data increases the resultant MSEs. This does not necessarily mean that using D2 will produce more accurate parameter estimates than using D3 as will be discussed in the next paragraph.

Table 5 contains the MAPEs between the true and estimated model parameters for case studies 2 and 3. The results show that the Adam-LBFGS optimiser approach for both cases produces smaller errors when compared to the Adam-only optimised parameter values. For case 2 utilising the Adam-LBFGS approach, the results achieved using only the arterial pressure and valvular flow data (D2) yielded smaller errors when compared to the results when using D3. Conversely, for case 3, the addition of heart chamber volume data (D3) resulted in the smallest MAPE. Therefore, using D2 for the aortic stenosis case and D3 for the mitral regurgitation case along with the Adam-LBFGS optimiser resulted in the best parameter estimates.

\begin{table}[htbp]
  \centering
  \caption{Mean absolute percentage errors for case studies 2 and 3 using datasets 2 and 3}
    \begin{tabular}{p{1.5cm} p{1cm} p{1cm} p{1cm} p{1cm} p{1cm} p{1cm}}
    \hline
          & \multicolumn{3}{c}{Case study 2} & \multicolumn{3}{c}{Case study 3} \\
          & Adam & Adam + LBFGS & FDM & Adam & Adam + LBFGS & FDM \\
    \hline
    Dataset 2 & 17.50 & 7.68  & 17.09 & 8.01  & 6.79  & 88.14 \\
    Dataset 3 & 18.95 & 14.14 & 21.03 & 7.05  & 6.16  & 76.17 \\
    \hline
    \end{tabular}
  \label{case23_mape_table}
\end{table}

\begin{figure}[h!]
	\centering
	\includegraphics[width=\textwidth]{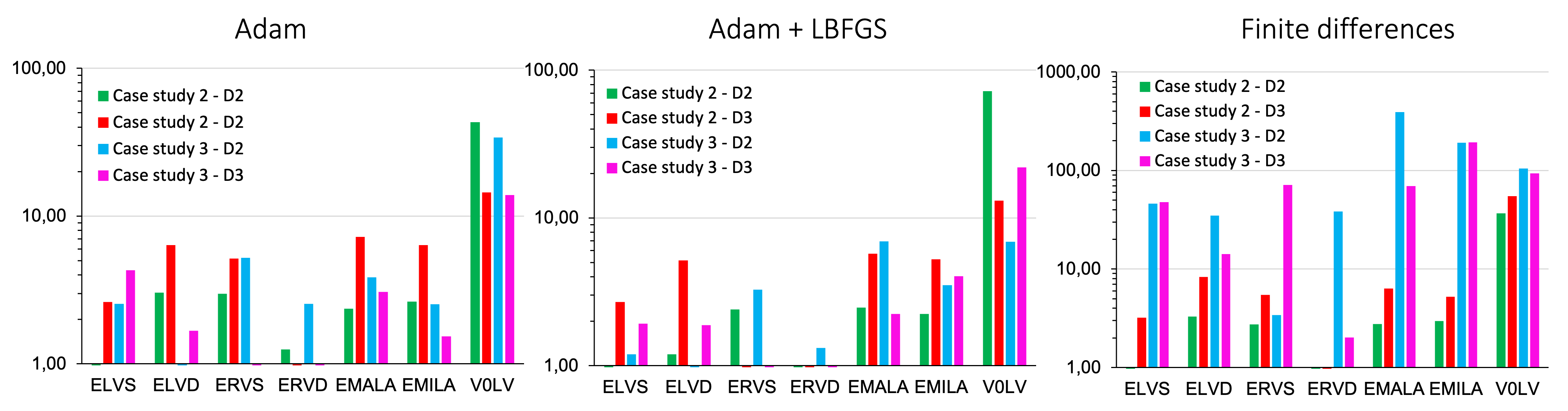}
	\caption{Heart parameter absolute percentage errors for case studies 2 and 3 (X-axis legend see figure \ref{sens_res})}
	\label{histograms}
\end{figure}

To further investigate the effects of measurement data and valvular diseases on the heart parameter estimates, bar plots of the individual APEs are plotted in figure \ref{histograms} for the different optimisation approaches. The results show that the FDM struggles to accurately estimate the model parameters for case study 3, and performs similar to the Adam optimised model for case study 2 which aligns with training history observed in figure \ref{case23_hists}. Studying the Adam and Adam-LBFGS error plots for both case studies, it is seen that using D2 for model training, results in smaller LV elastance parameter errors when compared to the error results generated when using D3, but generally using D3 for model training yields smaller errors for the remaining parameters shown in figure \ref{histograms} along with the various vasculature parameters (not shown). Therefore, we can conclude that the addition of the volume waveform information into the measurement dataset does not produce more accurate LV elastance parameter estimates but does increase the accuracy of the remaining parameter estimates at the cost of a slight increase in LV elastance parameter errors.  

\begin{figure}[h!]
	\centering
	\includegraphics[width=\textwidth]{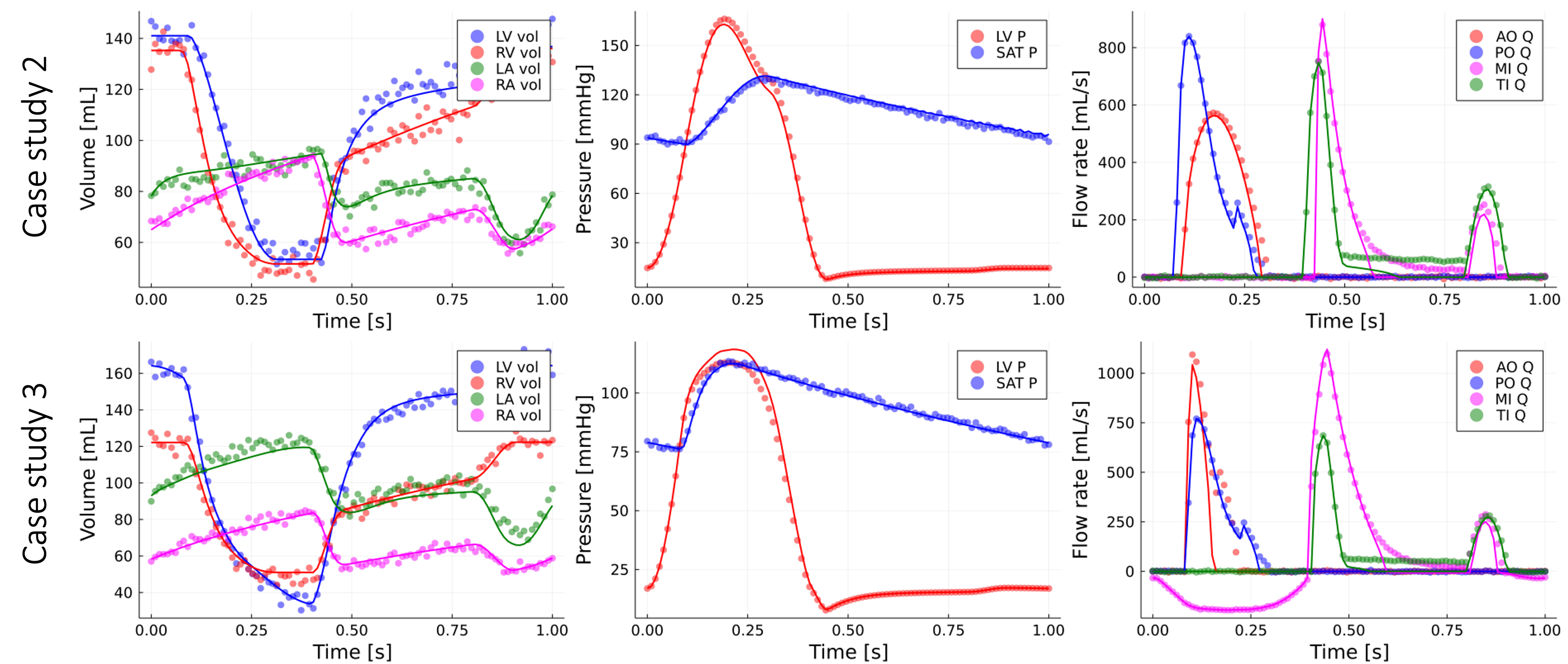}
	\caption{Predicted heart chamber volumes, arterial pressure and valvular flow rates using dataset 2 (markers – synthetic measurement data, line – simulated using optimised parameters)}
	\label{case23_d2}
\end{figure}

\begin{figure}[h!]
	\centering
	\includegraphics[width=\textwidth]{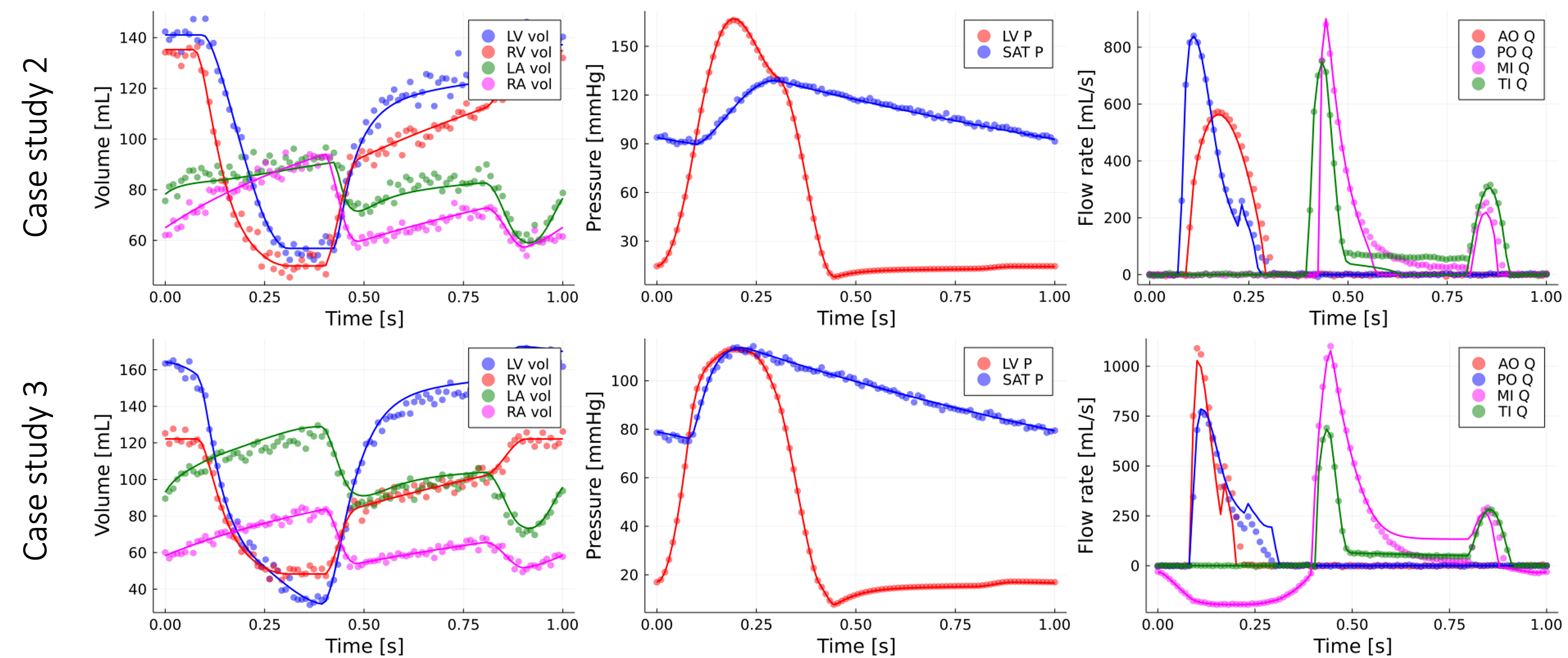}
	\caption{Predicted heart chamber volumes, arterial pressure and valvular flow rates using dataset 3 (markers – synthetic measurement data, line – simulated using optimised parameters)}
	\label{case23_d3}
\end{figure}

To visualise the effect of using the different measurement datasets, the predicted and actual volume, pressure and flow rate wave forms are plotted in figures \ref{case23_d2} and \ref{case23_d3} for parameter sets estimated using D2 and D3 with the Adam-LBFGS optimiser. It is seen in figure \ref{case23_d2} that the predicted parameters can clearly replicate the heart chamber volume waveforms, along with the arterial pressure and valvular flow rates for both case studies. For the unobserved LV pressure it is seen that the model slightly under predicts the pressure for case 2 and slightly overpredicts the LV pressure for case 3. 

In figure \ref{case23_d3}, similar to the results in figure \ref{case23_d2}, it is seen that the LPM with the optimised parameters using D3 can with relative accuracy replicate the actual waveforms. It is seen that the predicted LV pressure accurately tracks the actual LV pressure, and in contrast to the results in figure \ref{case23_d2}, does not over- or under-shoot the true waveform. This can be ascribed to the fact that the addition of heart chamber volume waveforms in the measurement data, leads to the optimiser more accurately capturing the various compartment resistances, capacitances and blood inertias which make up the ventricular afterload. Therefore, if the equipment is available to accurately measure the atrial and ventricular volume changes it is recommended to use the volume data in addition to the flow rate and pressure data when optimising the model parameters to accurately capture the ventricular afterload. That being said, using only the flow rate and pressure data, based on the current findings, would still yield accurate elastance parameter estimates. 

\section{Limitations}
In the present work, it is assumed that the lumped parameter/0D modelling approach is sufficient to capture the dynamics of an actual cardiovascular system. Furthermore, the heart valve flow coefficients were kept at constant values taken from literature, but in reality these flow coefficients would change based on the varying flow Reynolds number in the valve. Another limitation of the current work, is that the sensitivity analysis results are dependent on the selected parameter ranges, which in the current work is not based on values found in literature and rather selected to cover a wide range of vasculature impedances and heart chamber elastances.

\section{Conclusions}
The goal of the present work was to demonstrate the ability of computational parameter estimation algorithms to estimate LV elastances for diseased heart valve cases. It was shown using synthetically generated measurement data, that the use of automatic differentiation over finite differences to calculate the required gradients for cardiovascular parameter estimation, significantly improves the overall accuracy of the optimised parameters as well as the computational efficiency. Various gradient-based optimisers were also evaluated and it was found that a combination of Adam and LBFGS optimisers produced the best results for the analysed configurations. Using the hybrid Adam-LBFGS optimiser, important parameters related to LV performance was estimated for cases with aortic stenosis and mitral regurgitation using different collections of measurements. It was found that using arterial pressure and valvular flow rate waveforms along with the proposed optimiser, the approach was capable of predicting the LV elastances within approximately 2\% of the actual values. Adding the ventricular and atrial volume waveforms to the training data increased the LV elastance value errors to approximately 5\%, but the additional information meant the model was capable of more accurately capturing the various vasculature parameters. The combination of slightly higher LV elastance and lower vasculature parameter errors, resulted in simulation results that more closely track the actual LV pressure waveforms. Future work entails validating the proposed approach with actual patient data with known LV elastances.

\bibliographystyle{elsarticle-num}
\bibliography{Cardiovascularmodelling.bib}

\end{document}